# Studies on Dielectric Relaxation in Ceramic Multiferroic $Gd_{1-x}Y_xMnO_3$

R.M. Sarguna,[1*] V. Sridharan,[1] A.M. Awasthi,[2] and N. Subramanian[1]

[1]*Condensed Matter Physics Division, Materials Science Group,
Indira Gandhi Centre for Atomic Research, Kalpakkam- 603102, India*
[2]*Thermodynamics Laboratory, UGC-DAE Consortium for Scientific Research,
University Campus, Khandwa Road, Indore- 452001, India*

**ABSTRACT**

Dielectric study over a broadband was carried out from 10 to 70 K on ceramic $Gd_{1-x}Y_xMnO_3$ (*x*=0.2, 0.3, & 0.4). For all the compositions, a prominent sharp peak about ~18K was observed in the temperature dependence of both $\varepsilon'(T)$ and $\varepsilon''(T)$ at all frequencies, indicating a long-range ferroelectric (FE) transition. Using Cole-Cole fit to the permittivity data, the relaxation time $\tau$ and the dielectric strength $\Delta\varepsilon$ were estimated. Temperature variation of $\tau(T)$ in the Arrhenius representation is found to be non-linear (non-Debyean relaxation), with increasing barrier-activation energy over successive temperature-windows. Interestingly, for all the compositions, we witness a jump in $\tau(T)$ about the ferroelectric transition temperature, concurred by a broad-maximum in $\Delta\varepsilon(T)$, signifying the critical slow down of relaxations near long-range FE-correlations.

*rms@igcar.gov.in

**Introduction**

Materials exhibiting more than one type of ferroic property, magnetic, electric and strain in a single phase are called multiferroic[1]. Most of the works on multiferroics are devoted to materials exhibiting long range magnetic and electric ordering[2]. Presence of coupling between different ferroic properties in multiferroics offers additional advantage of controlling one property through another, especially modification of electric property through magnetic field. Thus, multiferroics have caught the attention of scientists both from technological and basic research point of view. Ortho-Perovskite manganites, $RMnO_3$ (R = rare-earth) belong to this class of materials and exhibit coupling between their magnetic and ferroelectric property[2]. Variation in the rare-earth ionic radii modifies the crystallographic structure through orthorhombic distortion, which in turn changes the magnetic structure of $RMnO_3$[3]. With decrease in rare-earth ionic radius $r_R$, the crystal symmetry changes from orthorhombic for La-Gd to hexagonal for Ho-Lu. The magnetic ground state of $Mn^{3+}$-sublattice changes from A-type (La-Gd) to E-type antiferromagnetic (Ho-Lu), via an incommensurate modulated magnetic structure for the intermediate (Tb-Dy) $RMnO_3$ systems. For the intermediate values of $r_R$, to which $TbMnO_3$ and $DyMnO_3$ belong, two magnetic transitions are reported; paramagnetic to incommensurate sinusoidal AFM structure at $T_{N1}(Mn^{3+})$ and from sinusoidal to cycloidal AFM structure at $T_{N2}(Mn^{3+})$[2,4,5]. The $RMnO_3$ systems falling in this mid-range of $r_R$ are intrinsically multiferroic; exhibiting an FE state. On the other hand, systems e.g., $GdMnO_3$ and $EuMnO_3$ are not intrinsically multiferroic[4,5].

Extensive dielectric studies have been carried out on $RMnO_3$ system and mixed crystals[5-9]. Strong anomaly in the temperature dependent dielectric constant $\varepsilon'(T)$ has been reported only about $T_{N2}(Mn^{3+})$. Onset of spontaneous ferroelectric polarisation in multiferroic $RMnO_3$ compounds like $TbMnO_3$ and $DyMnO_3$ is marked by a sharp peak in $\varepsilon'(T)$[4]. However, a step-like feature is observed about $T_{N2}(Mn^{3+})$ for the paraelectric like $GdMnO_3$. For the mixed crystal $(Tb_{1-x}Gd_xMnO_3)^6$, a sharp peak is observed for the concentration range up to $x \leq 0.8$, within which the system is multiferroic. Beyond this range, the system is paraelectric and the peak in $\varepsilon'(T)$ modifies to a step-like anomaly, quite similar to that reported for $GdMnO_3$. Likewise, external magnetic field along the *b*-axis renders $GdMnO_3$ multiferroicity and the corresponding anomaly modifies from a step to peak in $\varepsilon'(T)$. Thus, it is evident that a sharp peak rather than a step in $\varepsilon'(T)$ about $T_{N2}(Mn^{3+})$ signifies a ferroelectric transition.

Though substantial amount of dielectric studies have been reported, scant work has been carried out on the relaxation associated with this class of multiferroics. To the best of our knowledge, a lone report by Schrettle et. al.[10] exists on such studies. They found a critical slowdown of the relaxation time ($\tau$, evidencing the FE state), concurring below the spiral spin ordering transition in *multiferroic* manganites, in contrast to its step-acceleration at the *canted-*AFM type ordering in all the other manganites, followed by the freezing-in of the latter's dipolar relaxations. In view of this, it is of importance to study the dielectric relaxation behaviour in the class of strictly *multiferroic* manganites. In the present paper, we report the dielectric properties of $Gd_{1-x}Y_xMnO_3$ system from 10 to 70 K over a frequency range 100 Hz to 125 kHz. From Cole-Cole fitting, the relaxation time as the function of temperature was estimated after taking into account the low-frequency dominant conductivity part. Temperature dependence of kinetics thus evaluated illustrates the non-Arrhenic character of relaxation in this system as well as its critical slowdown at $T_{N2}(Mn^{3+})$, signifying the ferroelectric transition.

**Experimental Details**

The polycrystalline samples of $Gd_{1-x}Y_xMnO_3$ ($x$ = 0.2, 0.3, & 0.4) were synthesized by solid state synthesis route by heat treating stoichiometric mixture of 99.95% pure $Gd_2O_3$, $Y_2O_3$, and $MnO_2$ powders of Alfa-Aesar make. $Gd_2O_3$ was calcined over night at 900ºC to remove adsorbed moisture. Powder samples were heat treat in the temperature range 1000 to 1350 °C with two intermediate grinding. Finally, the powder was compacted in to pellets (8 mm diameter and ~ 1.5 mm thick) at 40 kN. Pellets were sintered at 1350 °C for 24 hrs. and furnace-cooled. Crushed fine powder was used to record the powder X-ray diffraction (XRD) patterns with Cu-K$_\alpha$ radiation, using a STOE diffractometer operated in the Bragg-Brentano geometry. Low temperature dielectric measurements were carried out in the temperature range of 5K to 60K using a broadband impedance analyzer (Alpha-A, NovoControl) over wide frequency range. The dielectric measurements were carried out on sintered pellets. Electrodes for dielectric measurements were made by applying silver paste on both sides of the pellets and annealed at 150 °C for one hour for curing of the silver paste. Typical two-probe resistance of the pellets across their faces was measured to be more than 20 MΩ and no detectable change in its value was observed after electrode formation.

**Results and Discussion**

The powder X-ray diffraction patterns of $Gd_{1-x}Y_xMnO_3$ ($x$=0.2, 0.3, & 0.4) system are shown in Fig.1. The pattern matches well with the JCPDS data of $GdMnO_3$ [ICDD No.01-070-9199]. The peaks in general shift to higher angle, indicating reduction in the interatomic distances. This is consistence as ionic radius of $Y^{3+}$ [1.075] is smaller than that of $Gd^{3+}$ [1.107]. All the peaks could be indexed to orthorhombic symmetry. The lattice parameters for all the composition are tabulated in Table 1. While the lattice parameters $a$ and $c$ decrease almost linearly, non-linear decrease in the lattice parameter $b$ is observed. However, the relation, $c/\sqrt{2} < a < b$ corresponding to O'-distortion of the ortho-perovskite is obeyed for all $x$. The orthorhombic distortion as parameterized by the tolerance factor (Table 1) decreases with increase of $x$, indicating increased distortion. The SEM image of the ceramic pellets is shown in Fig.2. The pellets though sintered at high temperature do exhibit porosity. As no change in the 2-probe face to face resistance before and after application of silver paste is observed, it is inferred that pores may be close-ended. Temperature variation of the real part of the dielectric constant, $\varepsilon'(T)$ of $Gd_{1-x}Y_xMnO_3$ ($x$ = 0.2, 0.3, and 0.4) for various frequencies is shown in Fig.3. For the sake of comparison, $\varepsilon'(T)$ measured at 10 kHz for all the compositions are shown in Fig.3(d). For all compositions and frequencies, an anomaly in the form of peak at ~18K and a weak shoulder at ~7 K are observed. The peak at 18 K downshifts nominally by ~2 K, as $x$ is increased from 0.2 to 0.4. However, there is no detectable change in the position of the shoulder with $x$. The width of the peak at 18K increases perceptibly with substitution, so that the shoulder (~7K) which is relatively well-resolved for $x$ = 0.2, is barely seen for $x$ = 0.4. In the parent compound $GdMnO_3$, a step-like anomaly instead of a peak at ~23K has been reported, associated to an incommensurate sinusoidal to an A-type antiferromagnetic transition[4,5]. On the other hand, a peak in $\varepsilon'(T)$ has been reported for the multiferroic systems $RMnO_3$ ($TbMnO_3$ and $DyMnO_3$) and $(RR')MnO_3$ ($Eu_{1-x}Y_xMnO_3$, $Tb_{1-x}Gd_xMnO_3$[6] and $Sm_{1-x}Y_xMnO_3$[8]). This peak is associated with improper ferroelectric transition induced by magnetic transition. Thus, it is inferred that Y-substituted $GdMnO_3$ results in ordered ferroelectric state, otherwise induced in the parent compound only under an external magnetic field. Accordingly, the peak is ascribed to metamagnetic transition of $Mn^{3+}$-sublattice ($T_{N2}$ ($Mn^{3+}$)). The peaks in the present study are not as sharp as reported for the $c$-axis oriented single crystals. This is attributed to polycrystalline nature of the sample and substitution of $Y^{3+}$ at $Gd^{3+}$, resulting in the site-disorder of the atoms, which may in turn lead to broadening of the peak associated with the ferroelectric transition. However, the site-disorder is not strong enough to alter the nature of the transition to the relaxor-type. This is clearly evident from the fact that the position of the peak in $\varepsilon'(T)$ is unaltered with frequency. Lower value of the dielectric permittivity may be attributed to the porosity present in the sample, as porosity is known to reduce the dielectric permittivity[11,12]. It is pertinent to note that though porosity affects the absolute value of the dielectric permittivity, the transition temperature is least affect by it.

Moderate decrease in the strength of the anomaly with frequency is noticed, and is in accordance with those reported for other $RMnO_3$ compounds. The shoulder observed in the real part of the dielectric constant about 7K is ascribed to paramagnetic to canted-AFM ordering of the $Gd^{3+}$ sublattice ($T_{N1}$ ($Gd^{3+}$))[4,5]. This is a clear indication that the rare-earth magnetic ordering also contributes to the dielectric permittivity, and has been reported for the parent compound $GdMnO_3$ as well[5]. Position of this peak is unaffected by both frequency and Y-substitution. However, no feature associated with the paramagnetic to sinusoidal AFM transition ($T_{N1}$ ($Mn^{3+}$)) of the $Mn^{3+}$ sublattice could be observed. This in general is in agreement with earlier works on the $RMnO_3$ systems[4].

Temperature dependent imaginary part of the dielectric constant $\varepsilon''(T)$ of $Gd_{1-x}Y_xMnO_3$ ($x$ = 0.2, 0.3, and 0.4) measured under different frequencies is shown in Fig.4. For ease of comparison, $\varepsilon''(T)$ measured at 10 kHz for $x$=0.2, 0.3, and 0.4 is also shown. The imaginary part of dielectric constant is strongly dependent on frequency and shows considerable dispersion. As in the case of $\varepsilon'(T)$, two anomalies at about 7 and 18 K are observed for $\varepsilon''(T)$. As had been inferred above, these anomalies are associated to relaxation processes related to $T_{N1}$ ($Gd^{3+}$) and

$T_{N2}$ ($Mn^{3+}$) respectively. Unlike $\varepsilon'(T)$, considerable suppression in the strength and dispersion in $\varepsilon''(T)$ about $T_{N2}$ ($Mn^{3+}$) is seen as $x$ is varied. This reduction renders the contribution of $Gd^{3+}$ sublattice to be distinct for $f \geq 10$ kHz.

Apart from these anomalies, additional ones emerge distinctly at still higher temperatures for $f \geq 3.78$ kHz. This peak evolves stronger and shifts to higher temperature with frequency, indicating a relaxation process[13]. This peak has been reported in the single crystal studies of $RMnO_3$ as well[4]. Though the peaks are resolved at higher frequencies, substantial overlap is observed at lower frequencies. Maximum in the peak position of $\varepsilon''(T)$ for the additional anomaly could be reliably estimated by fitting to the Debye model[14], given by

$$\varepsilon''(\omega, T) = \frac{\Delta\varepsilon}{2} \Big/ \cosh\left[\frac{U}{k_B}\left(\frac{1}{T_M} - \frac{1}{T}\right)\right] \qquad (1)$$

Where $\Delta\varepsilon (\equiv \varepsilon'_0 - \varepsilon'_\infty)$ is the dielectric strength, $U$ is the activation energy associated with the relaxation process, $k_B$ is the Boltzmann constant, and $T_M$ is the temperature where $\varepsilon''(T)$ is maximum. The $\varepsilon''(T)$ curves were fitted (Fig.5) with the above equation and $T_M$ was estimated. It is observed that the fitting is better for high frequencies, where there is strong overlap with the anomaly associated with $T_{N2}$ ($Mn^{3+}$).

Fig. 6 shows the Arrhenius plot ($\ln(\tau)$ vs. $1/T$) for all the compositions. It is clearly seen that the kinetics is non-Debyean. In order to have rough estimates of activation energy, the temperature variation of relaxation time were piecewise linearly fitted as shown in Fig. 6. The first slope change occurs near the $T_{N1}(Mn^{3+})$ temperature i.e., paramagnetic to sinusoidal antiferromagnetic transition of the $Mn^{3+}$ sublattice near about ~ 41 K for all the compositions. The activation energies estimated fall in the region of 22 meV to 8 meV, which are very well associated to electronic relaxation process. Similar trends and values for activation energy have been quoted for pure $GdMnO_3$[5].

The relaxation process related to ferroelectric phenomena can be best studied by analyzing the frequency ($\omega/2\pi$) dependence of dielectric permittivity at different temperatures. The loss peaks $\varepsilon''(\omega)$ over five decades of frequency for different temperatures across the ferroelectric transition are plotted in Fig. 7. The conductivity contribution is seen as an increase in the value of $\varepsilon''(\omega)$ at lower frequency below 100 Hz. This low-frequency dependence is given by[13]

$$\varepsilon'' = \sigma_0/\varepsilon_0\omega \qquad (2)$$

where $\sigma_0$ is the dc-conductivity of the sample and $\varepsilon_0$ is the dielectric permittivity of vacuum. With the decrease of temperature, the loss peaks shift towards lower frequencies with an increase in the strength. The loss peaks are much broader and indicate a distribution in the relaxation time. Hence the peaks are fitted using empirical Cole-Cole function, which takes this into account[13]. $\varepsilon''(\omega)$ was fitted to the Cole-Cole function for estimation of relaxation time.

$$\varepsilon''(\omega) = \left[\Delta\varepsilon * (\omega\tau_{cc})^\beta * \sin(\beta\pi/2)\right]/\left[1 + 2(\omega\tau_{cc})^\beta * \cos(\beta\pi/2) + (\omega\tau_{cc})^{2\beta}\right] \qquad (3)$$

Where $\Delta\varepsilon$ is the dielectric strength, $\tau_{cc}$ is the Cole-Cole relaxation time, and $\beta$ is a width parameter, with $\beta = 1$ implying the purely Debyean case. $\beta < 1$ implies broadening of the loss peak. As the $\varepsilon''(\omega)$ consists of conductive and relaxation contributions, the curves were fitted with both the conductivity contributions (given by Eq.2) and relaxation part given by a Cole-Cole function (Eq.3). The fitting for $Gd_{0.8}Y_{0.2}MnO_3$ at 23K is shown in Fig.8 as a representative.

Variation in the estimated relaxation time as a function of temperature is shown in Fig.9 in the Arrhenius representation, for all the compositions over the temperature range of 40K-10K, with the inset showing the full range down to 5K. The curves exhibit a non-linear behaviour. This essentially indicates that the relaxation process is not a single thermally-activated behaviour; in other words it is non-Debyean. The curvature in the $\log(\tau)$ vs. $1/T$ plots implies that a tunneling processes is involved at low temperatures[10].

A jump in the temperature variation of relaxation time $\tau(T)$, showing a slowdown of relaxation centered about the ferroelectric transition temperature is observed which is more evident for $x$=0.4 sample. This is followed by a broad peak (shown in the inset) near the $Gd^{3+}$-magnetic transition temperature, only to dramatically decrease at lower temperatures. The temperature variation of the dielectric strength $\Delta\varepsilon$ is plotted in Fig.10. $\Delta\varepsilon(T)$ for all the compositions is seen to increase near the $Gd^{3+}$-transition temperature and decrease well across the ferroelectric transition temperature. These observations show that both the ferroelectric transition as well as the $Gd^{3+}$-magnetic ordering both has considerable influence on the dielectric relaxation process. It is reflective of the shoulder like feature that is seen in the dielectric variation $\varepsilon'(T)$ near the $Gd^{3+}$ sublattice ordering temperature about 7K apart from the ferroelectric transition at 18K. The general trend of the variation of relaxation time near the ferroelectric transition temperature is in overall agreement with an earlier report on $RMnO_3$ systems[10] but with differences. While a sharp peak in $\tau(T)$ has been reported for the intrinsically multiferroic $TbMnO_3$ and $DyMnO_3$, a step-like feature (acceleration) has been observed for the systems wherein the multiferroicity is induced either by external magnetic field ($GdMnO_3$)[10]. Otherwise, no anomaly in $\tau(T)$ is observed for the other systems. However, in the present study a sharp jump is observed at 18K near the ferroelectric transition temperature indicaticating slowdown of the relaxation process instead of a sharp peak, followed by a peak near the $Gd^{3+}$ ordering temperature. The variation in the shape of the peak at the ferroelectric transition could be ascribed to the closely following relaxation associated with $Gd^{3+}$-sublattice ordering as well as due to the polycrystalline nature of the samples in contrast to use of *c*-axis oriented single crystals by Schrettleet. et. al.[10]. Notwithstanding these differences, peak-like anomaly in both $\tau(T)$ and $\Delta\varepsilon(T)$ clearly signify slowing down of the relaxation process near the ferroelectric transition in $Gd_{1-x}Y_xMnO_3$.

**Conclusions**

Dielectric measurements were carried out on $Gd_{1-x}Y_xMnO_3$ ($x$=0.2, 0.3, and 0.4) from 10 to 70 K over a frequency range 100 Hz to 125 kHz. For all the compositions, an anomaly about 18K in the form of peak has been observed in the temperature variations of $\varepsilon'$ and $\varepsilon''$. This clearly implies that $Y^{3+}$ substitution endows multiferroicity in $GdMnO_3$, which is otherwise induced by an external magnetic field. The frequency variation of $\varepsilon''$ was fitted to the Cole-Cole function and the temperature variation of the relaxation time $\tau(T)$ and dielectric strength $\Delta\varepsilon(T)$ were estimated. Variation of $\tau(T)$ in Arrhenius representation is found to be non-linear with a sudden slow down of the relaxation process near the ferroelectric transition; features witnessed only in multiferroic materials[10]. Further, the presence of such anomalies in $\tau(T)$ and $\Delta\varepsilon(T)$ both in single and polycrystalline samples establishes the intrinsic nature of these features to multiferroic $RMnO_3$ systems.

**Figure Captions**

**Figure 1:** Room temperature XRD patterns of $Gd_{1-x}Y_xMnO_3$ ($x$=0.2, 0.3 and 0.4). The diffraction patterns are vertically shifted for clarity (only the major peaks have been indexed).

**Figure 2:** SEM micrographs of $Gd_{1-x}Y_xMnO_3$ for a) $x$=0.2, b) $x$=0.3, and c) $x$=0.4.

**Figure 3:** Variation of real part of dielectric permittivity with temperature $\varepsilon'(T)$ for $Gd_{1-x}Y_xMnO_3$: a) $x$=0.2, b) $x$=0.3, and c) $x$=0.4 for various frequencies, d) comparison of $\varepsilon'(T)$ measured at 10 kHz for $x$=0.2, 0.3, and 0.4. Vertical lines below ~10K and above ~40K correspond to magnetic transition temperatures of $Gd^3$ and $Mn^{3+}$ sublattices, respectively.

**Figure 4:** Variation of imaginary part of the dielectric permittivity $\varepsilon''(T)$ with temperature for $Gd_{1-x}Y_xMnO_3$: a) $x$=0.2, b) $x$=0.3, c) $x$=0.4 for various frequencies, and d) comparison of $\varepsilon''(T)$ measured at 10 kHz for $x$=0.2, 0.3, and 0.4. Vertical lines below ~10K and above ~40K correspond to magnetic transition temperatures of $Gd^{3+}$ and $Mn^{3+}$ sublattices, respectively.

**Figure 5:** Variation of $\varepsilon''$ with temperature at different frequencies. Curve fitting to estimate additional anomaly in $\varepsilon''(T)$ of $Gd_{0.7}Y_{0.3}MnO_3$ carried out with the Debye model.

**Figure 6:** Variation of relaxation time $\{\ln(\tau)$ vs. $(1/T)\}$ for a) $x$=0.2, b) $x$=0.3, and c) $x$=0.4; each revealing three different "Arrhenic" regimes. Corresponding activation energies $E_a$ in three distinct temperature regimes for each $x$ are as per indicated.

**Figure 7:** Figure showing the variation of $\varepsilon''(\omega)$ at different temperatures for $Gd_{0.8}Y_{0.2}MnO_3$.

**Figure 8:** Figure showing the fitting $\varepsilon''(\omega)$ with both conductive and Cole-Cole function for $x$= 0.2 at 23K.

**Figure 9:** Arrhenius representation of relaxation time variation as a function of temperature for $Gd_{1-x}Y_xMnO_3$ for ($x$=0.2, 0.3, and 0.4). Inset showing the same curves on linear scale over the full range.

**Figure 10:** Variation of dielectric strength $\Delta\varepsilon(T)$ for $Gd_{1-x}Y_xMnO_3$ ($x$ = 0.2, 0.3, & 0.4).

**Table 1:** Table showing the lattice parameters and tolerance factor for $Gd_{1-x}Y_xMnO_3$ ($x = 0.2$, 0.3, & 0.4).

| Composition | $a$ (Å) | $b$ (Å) | $c$ (Å) | Tolerance factor |
|---|---|---|---|---|
| $Gd_{0.8}Y_{0.2}MnO_3$ | 5.3048 (3) | 5.8536 (4) | 7.4190 (4) | 0.86462 |
| $Gd_{0.7}Y_{0.3}MnO_3$ | 5.2988 (1) | 5.8532 (1) | 7.4105 (1) | 0.8635 |
| $Gd_{0.6}Y_{0.4}MnO_3$ | 5.2927 (2) | 5.8511 (3) | 7.4027 (3) | 0.8624 |

**Figure 1**

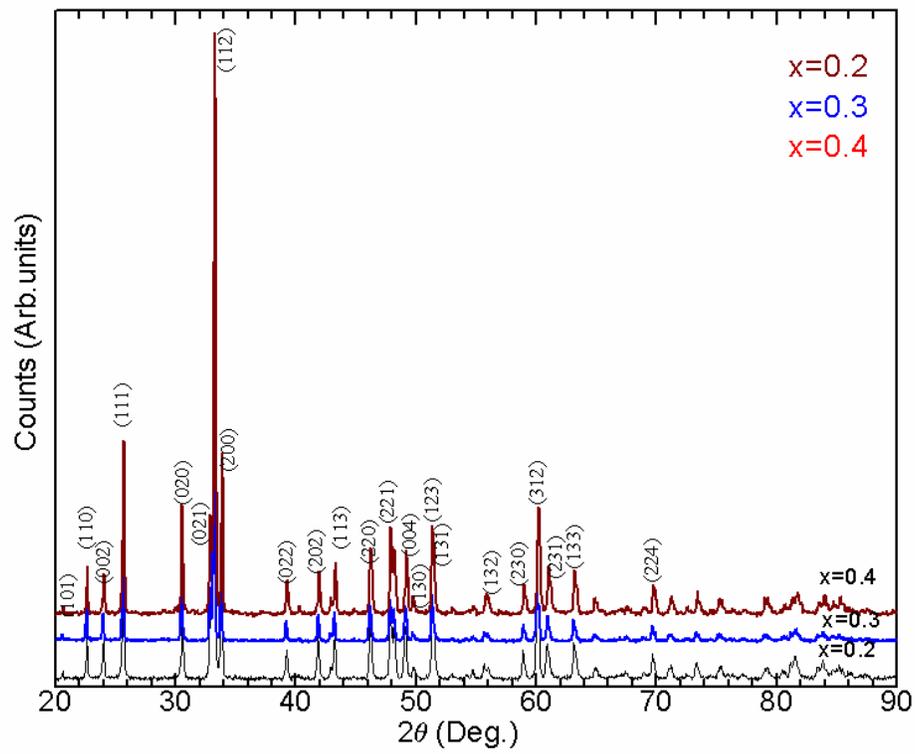

**Figure 2**

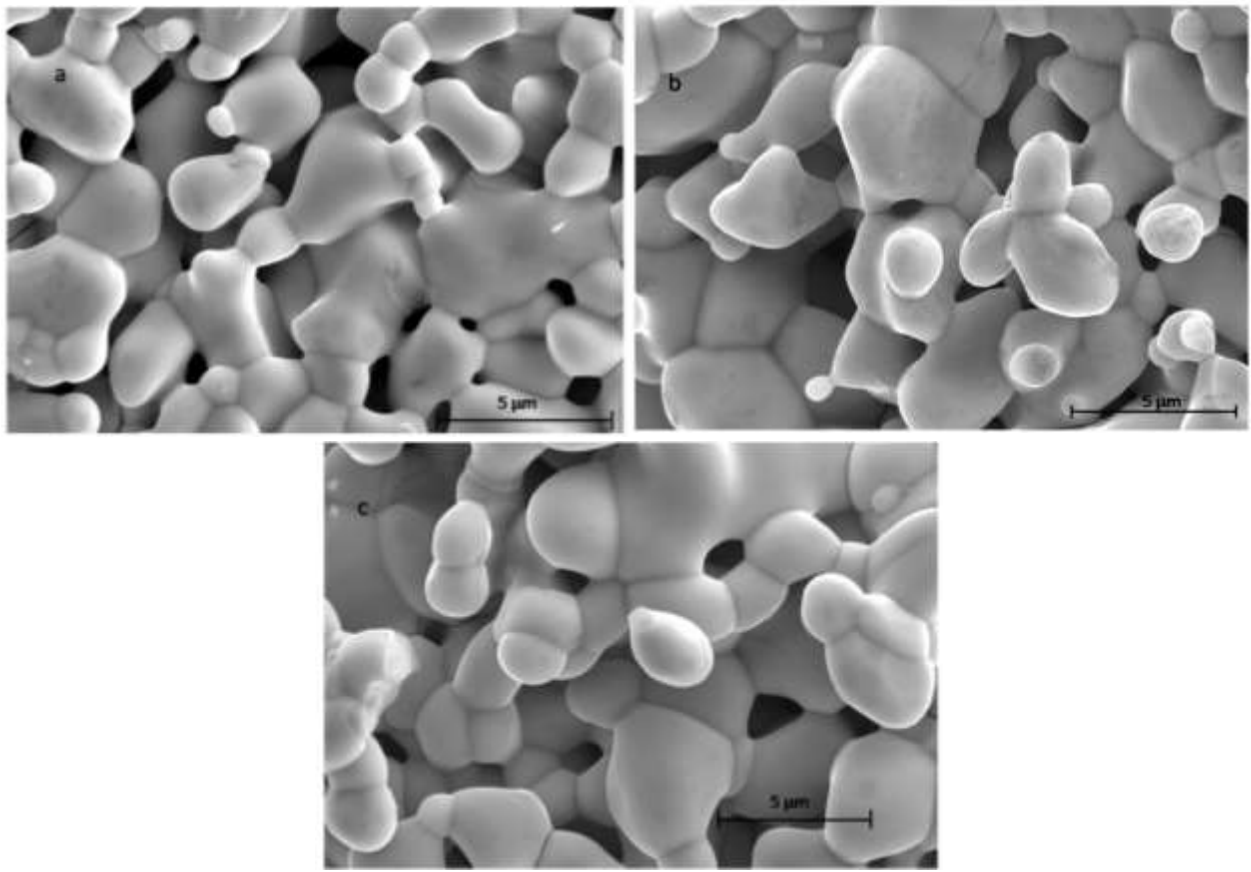

**Figure 3**

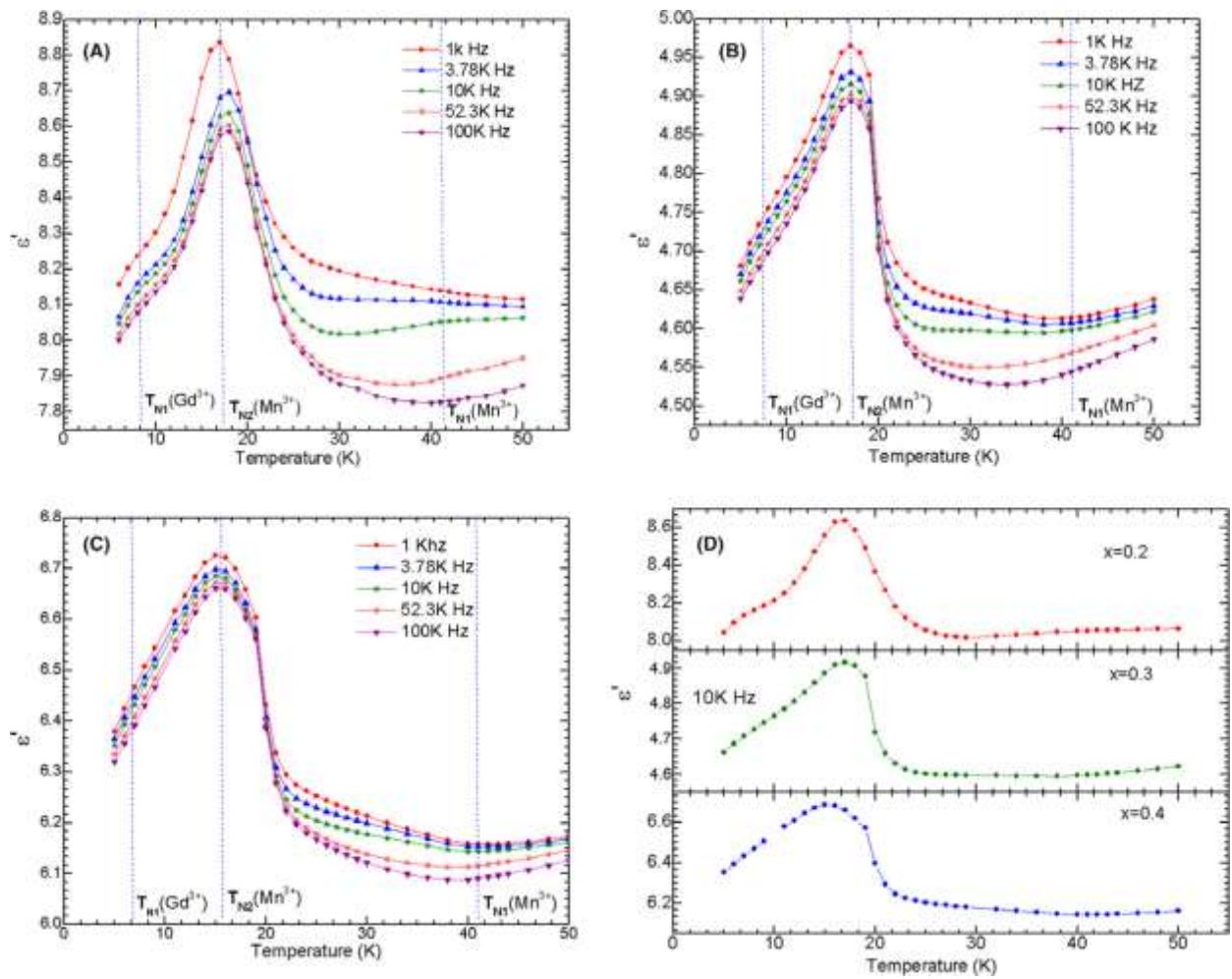

**Figure 4**

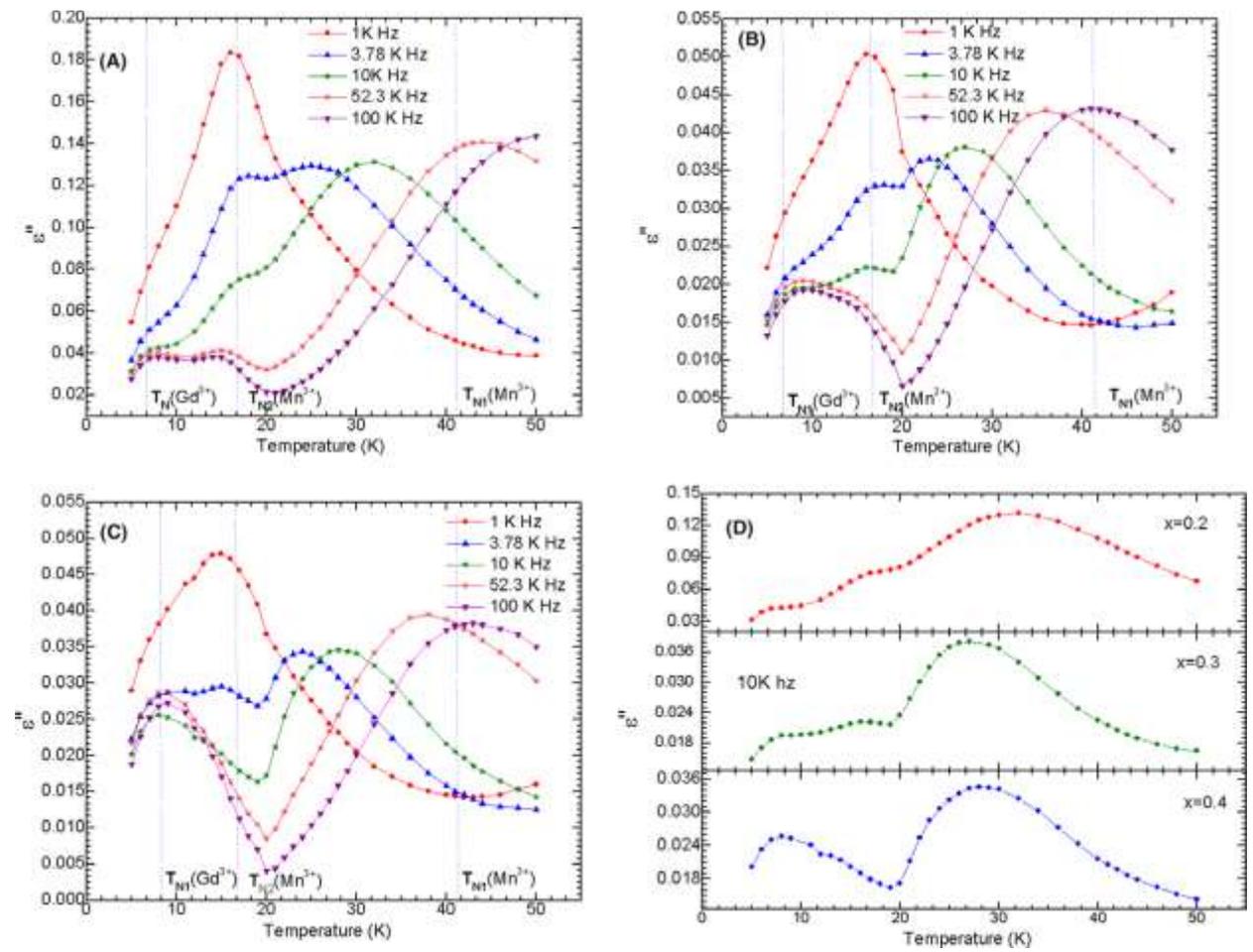

**Figure 5**

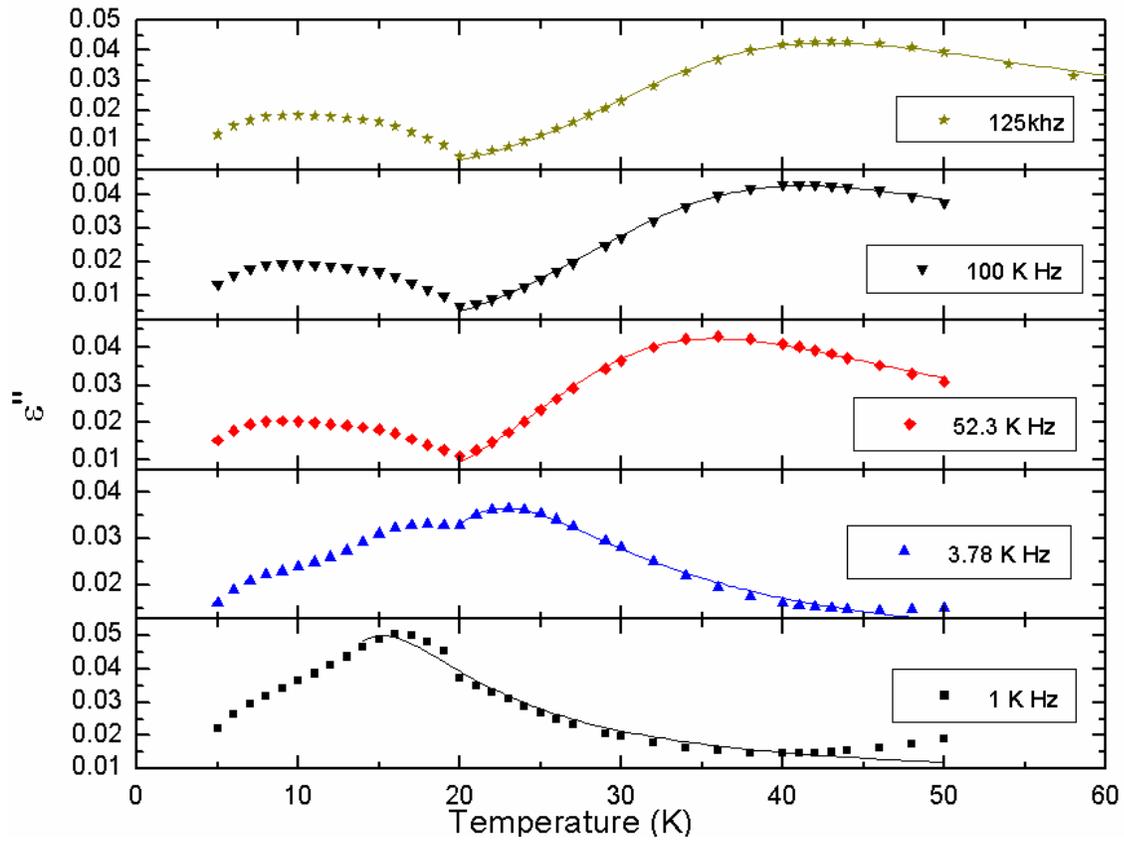

**Figure 6**

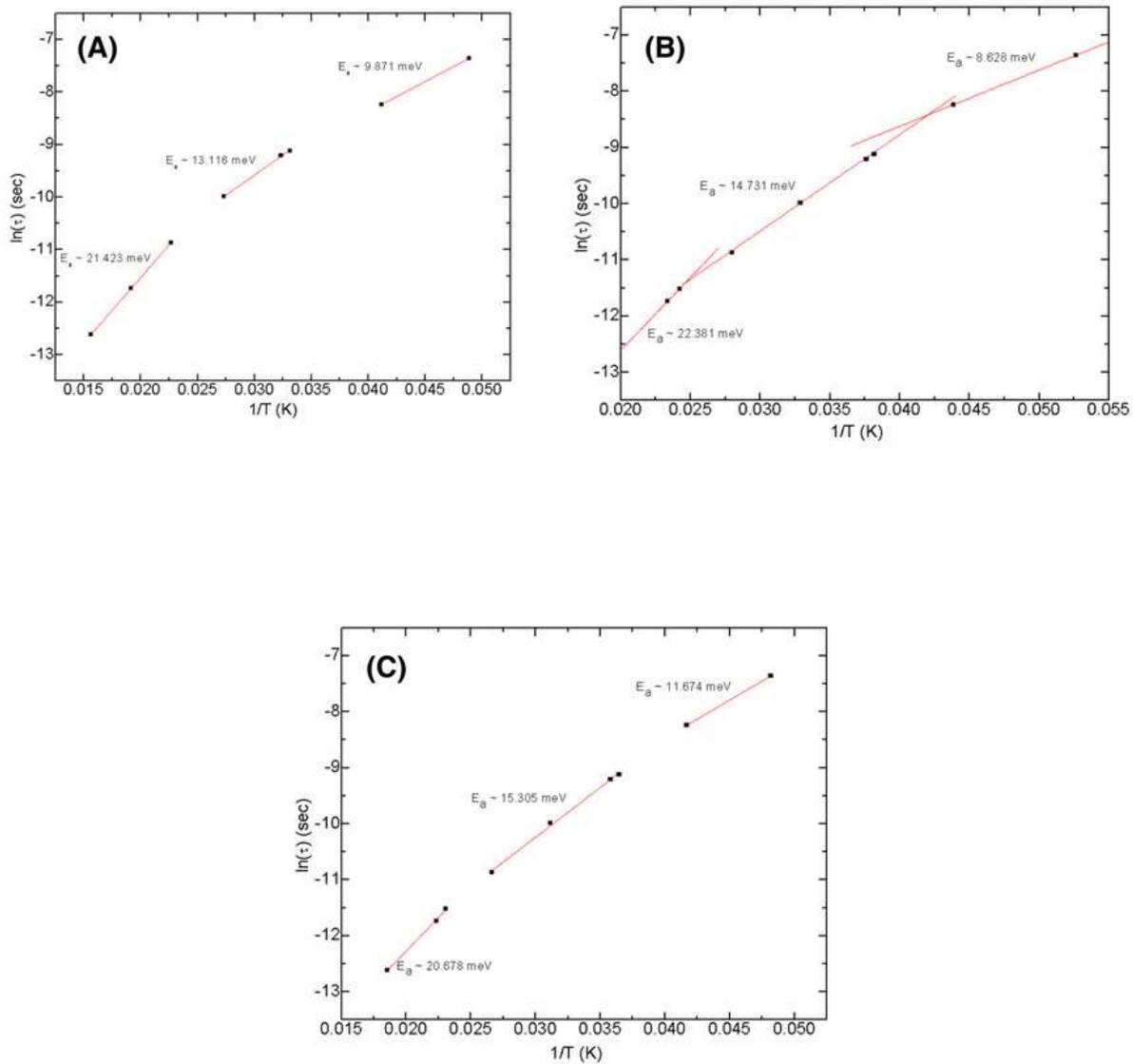

**Figure 7**

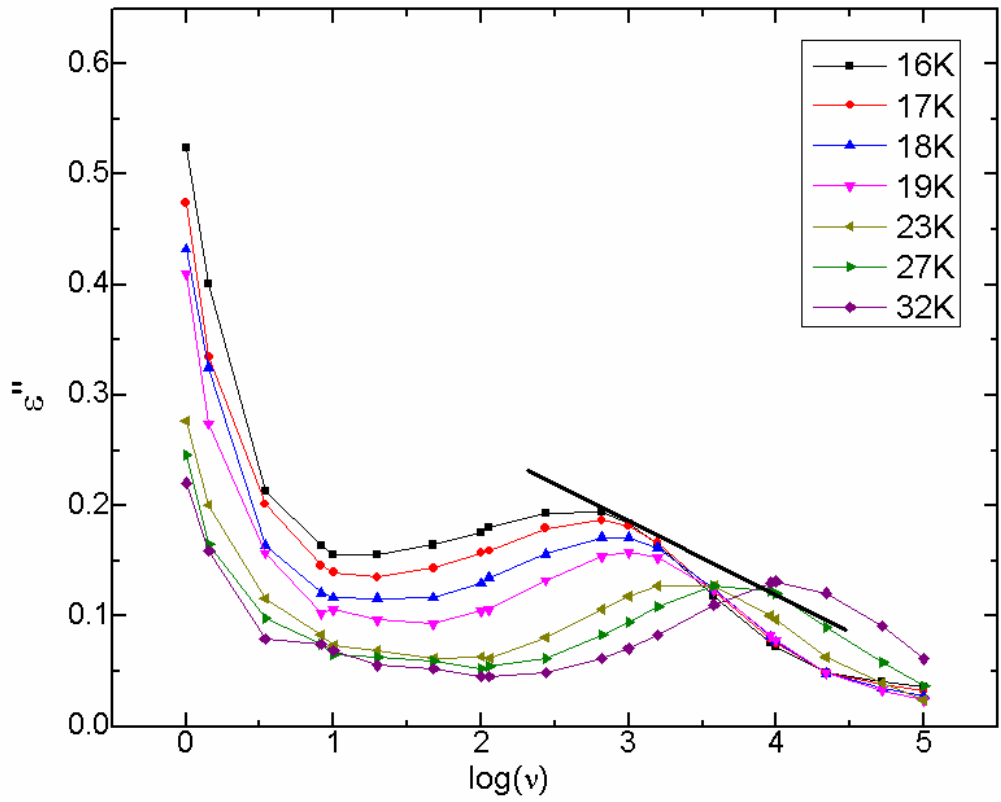

**Figure 8**

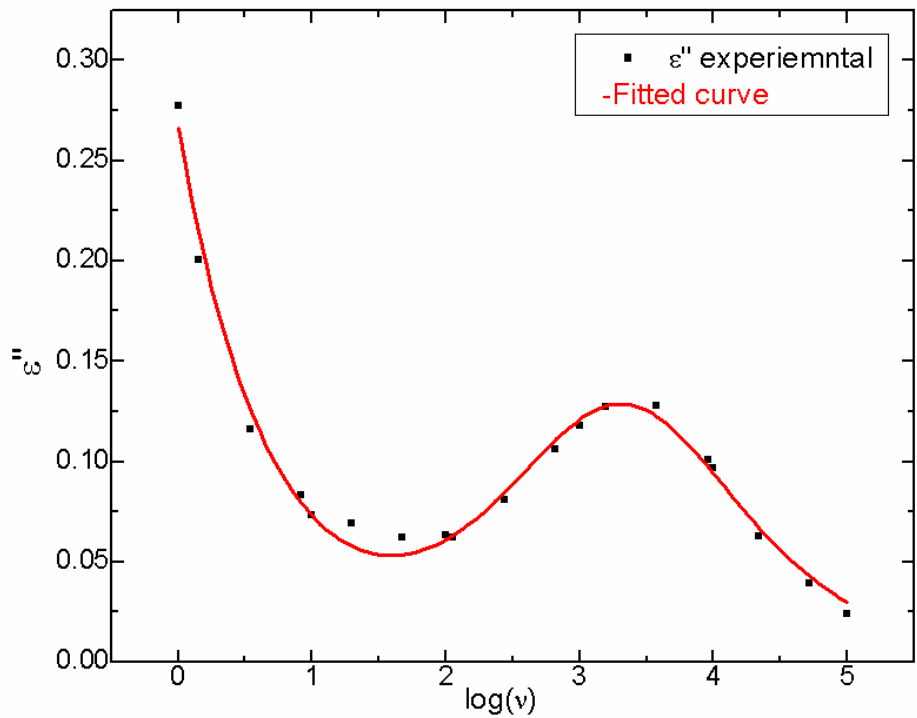

**Figure 9**

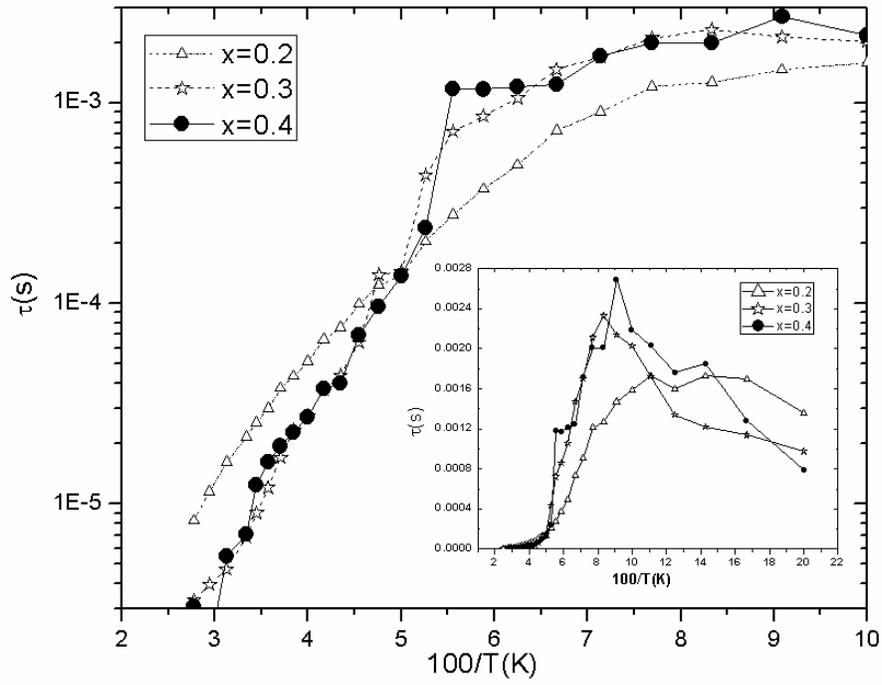

**Figure 10**

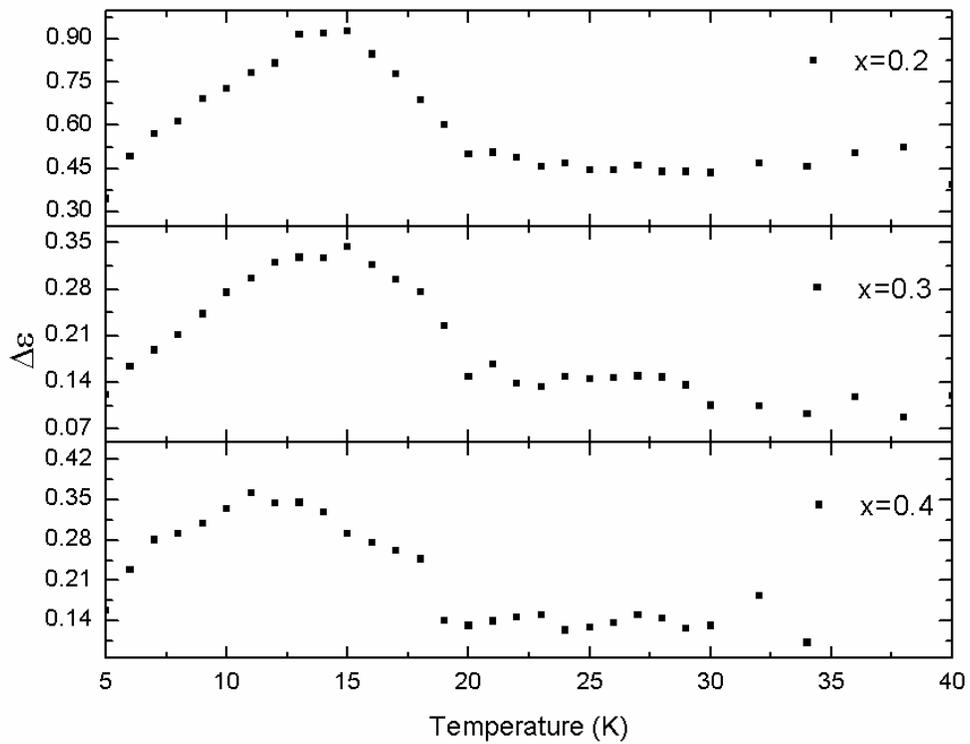